\documentclass[pra,aps,showpacs,twocolumn,superscriptaddress]{revtex4}
\usepackage{amsmath,amsfonts,amssymb,mathrsfs,bm}
\usepackage{graphicx,graphics,color,epsfig}
\begin{document}
[Phys. Rev. A {\bf 83}, 025601 (2011)]
\title{Orbital elementary excitations as probes of entanglement and quantum phase transitions of collective spins in an entangled Bose-Einstein condensate}
\author{Rukuan Wu}
\affiliation{State Key Laboratory of Surface Physics and  Department of Physics, Fudan University, Shanghai
200433, China}
\affiliation{Department of Physics, Zhejiang Normal
University, Jinhua 321004, China}
\author{Yu Shi}
\email{yushi@fudan.edu.cn}
\affiliation{State Key Laboratory of Surface Physics and  Department of Physics, Fudan
University, Shanghai 200433, China}
\begin{abstract}

A mixture of two species of pseudospin-$\frac{1}{2}$ Bose gases exhibits interesting interplay between spin and orbital degrees of freedom. Expectation values of various quantities of the collective spins of the two species play crucial roles in the Gross-Pitaevskii-like equations governing the four orbital wave functions in which Bose-Einstein condensation occurs. Consequently, the elementary excitations of these orbital wave functions reflect properties of the collective spins. When the coupling between the two collective spins is isotropic,  the energy gap of the gapped orbital excitation peaks, while there is a quantum phase transition in the ground state of the effective Hamiltonian of the two collective spins, which have previously been found to be maximally entangled.
\end{abstract}


\maketitle

\section{Introduction}

A current trend in condensed matter physics is the interplay between spin and orbital degrees of freedom. It is interesting to explore this topic in the realms of Bose-Einstein condensation (BEC), as physical effects on single-particle levels are often enhanced by Bose statistics. A possible avenue  is  spinor Bose gases \cite{leggett,pethick,pitaevskii,leggett2}. The perspective is more broadened if we consider a mixture of two distinct species of spinor Bose gases, where there are several coupling strengths for the collective spins of the two species, which depend upon the orbital wave functions. This situation leads to coupling between spin and orbital degrees of freedom.  For simplicity, let us focus on
a mixture of two species of pseudospin-$\frac{1}{2}$ Bose gases with both intraspecies and interspecies spin-exchange interactions~\cite{shi,shi2,shi3}. Interspecies spin-exchange causes interspecies entanglement and the so-called entangled Bose-Einstein condensation (EBEC)~\cite{shi0},  which is different from a mixture of two species of spinless atoms, where the two species are disentangled and individually undergoes BEC.  EBEC amplifies quantum entanglement from individual particles to macroscopic condensates. This ground state bears some analogies with the lowest energy state of a single species of pseudospin-$\frac{1}{2}$ atoms in a double well~\cite{al} or occupying two orbital modes~\cite{leggett2,ks,ale}, but there are also differences due to the fact that two atoms of distinct species are distinguishable and thus the numbers of atoms are respectively conserved.

It has been shown that the interspecies entanglement in the ground state is maximal at the  isotropic  parameter point of the effective Heisenberg coupling of the two collective spins, and that the larger the particle numbers of the two species, the steeper the entanglement peak~\cite{shi}.
In this brief report, after elucidating that the entanglement peak is indeed located at a quantum phase transition point, we study how the ground state of the collective spins affects the elementary excitations of the orbital wave functions in which EBEC occurs.  Furthermore, we  find that in the vicinity of this quantum phase transition, the energy gap of the gapped orbital elementary excitation is strikingly different from those of disentangled ground states. Away from the quantum phase transition point, the elementary excitations tend to approach those of a disentangled ground state.

\section{The Model}

Consider a dilute gas of two distinct species of atoms. Each atom has an internal degree of freedom represented as a pseudospin-$\frac{1}{2}$,
with $z$-component basis states $\uparrow$ and $\downarrow$. For a single species of pseudospin-$\frac{1}{2}$ gas,  it was argued that the conservation of a small total spin in the cooling process invalidates the single orbital mode approximation~\cite{ks,ale}. In our system, in contrast, under the single orbital mode approximation, the total spin of the mixture can still be arbitrarily small due to the distinguishability of the two species. Therefore, there is no reason against the usual single orbital mode approximation, which works well in most of the BEC systems. Moreover, a field theoretical approach without using single orbital mode approximation confirms its validation~\cite{ge}. Therefore,  for
each atom of species $\alpha (\alpha = a,b)$ and pseudospin
$\sigma$ ($\sigma=\uparrow,\downarrow$), we can safely assume that only the single-particle orbital ground state $\phi_{\alpha\sigma}(\textbf{r})$ is occupied. Then the many-body Hamiltonian can be written as
$
\mathcal{H}=\sum_{\alpha,\sigma}
f_{\alpha\sigma}N_{i\sigma}+\frac{1}{2}\sum_{\alpha,\sigma\sigma'}
K^{\alpha\alpha}_{\sigma\sigma'}N_{\alpha\sigma}N_{\alpha\sigma'}
+\sum_{\sigma\sigma'}
K^{ab}_{\sigma\sigma'}N_{a\sigma}N_{b\sigma'}
+K_e (a^\dagger_\uparrow
a_\downarrow b^\dagger_\downarrow b_\uparrow+a^\dagger_\downarrow
a_\uparrow b^\dagger_\uparrow b_\downarrow),$
where $\alpha_{\sigma}$ denoted annihilation operator associated with $\phi_{\alpha\sigma}(\textbf{r})$ of species $\alpha$,  $N_{\alpha\sigma}=\alpha^{\dagger}_{\sigma}\alpha_{\sigma}$ is the corresponding  number of atoms \cite{shi}.  For each specie $\alpha$, the total particle
number  $N_{\alpha}=N_{\alpha\uparrow}+N_{\alpha\downarrow}$ is a constant.  The coefficients $K's$ are shorthand for
$K^{\alpha\beta}_{\sigma_{1}\sigma_{2}\sigma_{3}\sigma_{4}}\equiv
g^{\alpha\beta}_{\sigma_{1}\sigma_{2}\sigma_{3}\sigma_{4}}
\int\phi^{*}_{\alpha\sigma_{1}}\phi^{*}_{\beta\sigma_{2}}\phi_{\beta\sigma_{3}}
\phi_{\alpha\sigma_{4}}d^{3}r$,
where  $g^{\alpha\beta}_{\sigma_{1}\sigma_{2}\sigma_{3}\sigma_{4}} \equiv 2\pi\hbar^{2}\xi^{\alpha\beta}_{\sigma_{1}\sigma_{2}\sigma_{3}\sigma_{4}}
/\mu_{\alpha\beta}$, with  $\xi^{\alpha\beta}_{\sigma_{1}\sigma_{2}\sigma_{3}\sigma_{4}}$ being the
scattering length for the scattering in which an $\alpha$-atom flips its pseudospin from
$\sigma_{4}$ to $\sigma_{1}$ while an $\beta$-atom flips its pseudospin from
$\sigma_{3}$ to $\sigma_{2}$, $\mu_{\alpha\beta}= m_{\alpha}m_{\beta}/(m_{\alpha} +m_{\beta})$ is the reduced mass.  For intraspecies scattering,
$K^{\alpha\alpha}_{\sigma\sigma}\equiv K^{\alpha\alpha}_{\sigma\sigma\sigma\sigma},
K^{\alpha\alpha}_{\sigma\bar{\sigma}}\equiv2K^{\alpha\alpha}_{\sigma\bar{\sigma}\bar{\sigma}\sigma}=2K^{\alpha\alpha}_{\bar{\sigma}\sigma\sigma\bar{\sigma}}$
for $\sigma\neq\bar{\sigma}$. For interspecies scattering,
$K^{ab}_{\sigma\sigma'}\equiv K^{ab}_{\sigma\sigma'\sigma'\sigma}$,
$K_{e}\equiv
K^{ab}_{\uparrow\downarrow\uparrow\downarrow}=
K^{ab}_{\downarrow\uparrow\downarrow\uparrow}$~\cite{note}.
$f_{\alpha\sigma}\equiv
\epsilon_{\alpha\sigma}-K^{\alpha\alpha}_{\sigma\sigma}/2$, where
$\epsilon_{\alpha\sigma}=
\int\phi^{*}_{\alpha\sigma}(-\hbar^2\nabla^2_{\alpha}/2m_{\alpha}
+U_{\alpha\sigma})\phi_{\alpha\sigma}d^{3}r$ is the single-particle energy.

For simplicity, we assume that the scattering lengths satisfy
$g^{\alpha\alpha}_{\sigma\sigma\sigma\sigma}=
g^{\alpha\alpha}_{\sigma\bar{\sigma}\bar{\sigma}\sigma}=
g_{\alpha},$
$g^{ab}_{\sigma\sigma\sigma\sigma}=g_{s}$, $g^{ab}_{\sigma\bar{\sigma}\bar{\sigma}\sigma}=g_{d}$,
$g^{ab}_{\uparrow\downarrow\uparrow\downarrow}=
g^{ab}_{\downarrow\uparrow\downarrow\uparrow}=g_{e}$. Moreover, we focus on the uniform  case $\phi_{\alpha\sigma}=1/\sqrt{\Omega}$, where $\Omega$ is the volume of the system.
The total spin operator of species $\alpha$ is $\mathbf{S}_{\alpha}=
\alpha^\dagger_\sigma\mathbf{s}_{\sigma\sigma'}\alpha_{\sigma'}$, where $\mathbf{s}_{\sigma\sigma'}$ is the single spin operator. In terms of  $\mathbf{S}_{\alpha}$, the
Hamiltonian can be transformed into that of two coupled giant spins  $S_{a}=N_{a}/2$ and $S_{b}=N_{b}/2$,
\begin{equation}
{\cal H}=J_{\perp}(S_{ax}S_{bx}+S_{ay}S_{by})+
J_z S_{az}S_{bz}+E_{0} \label{spinhamiltonian}
\end{equation}
where $J_{\perp} \equiv \frac{2g_{e}}{\Omega}$, $J_z \equiv \frac{2(g_{s}-g_{d})}{\Omega}$, $E_{0}=\frac{N_{a}(N_{a}-1)}{2\Omega}g_{a}+
\frac{N_{b}(N_{b}-1)}{2\Omega}g_{b}+\frac{N_{a}N_{b}}{2\Omega}(g_{s}+g_{d})$
is a constant.  Without loss of generality, let $S_{a} \geq S_{b}$. We focus on antiferromagnetic couplings $g_e >0$ and  $g_{s}-g_{d}>0$.

\section{quantum phase transition}

There is a quantum phase transition at the parameter point $g_{e}=g_{s}-g_{d}$. The ground states are qualitatively different in the limits of  $g_{e} \gg g_{s}-g_{d}$ and   $g_{e} \ll g_{s}-g_{d}$.

In a mean field approximation, the ground state is disentangled between the two species, and can be written as
$|G\rangle_{MF}=(e^{-i\varphi_{a}/2}{\rm cos}\frac{\theta_{a}}{2}|\uparrow\rangle_{a} +e^{i\varphi_{a}/2}{\rm
sin}\frac{\theta_{a}}{2}|\downarrow\rangle_{a})^{N_{a}}\otimes (e^{-i\varphi_{b}/2}{\rm
cos}\frac{\theta_{b}}{2}|\uparrow\rangle_{b} +e^{i\varphi_{b}/2}{\rm
sin}\frac{\theta_{b}}{2}|\downarrow\rangle_{b})^{N_{b}},$  
with mean-field energy
$E_{MF}=\frac{2g_{e}}{\Omega}(\langle S_{ax}\rangle\langle
S_{bx}\rangle+\langle S_{ay}\rangle\langle S_{by})\rangle)
+\frac{2(g_{s}-g_{d})}{\Omega}\langle S_{az}\rangle\langle
S_{bz}\rangle,$
where a constant is neglected, $\langle S_{\alpha x}\rangle=\frac{N_{\alpha}}{2}{\rm sin}\theta_{\alpha}{\rm
cos}\varphi_{\alpha},$ $\langle S_{\alpha y}\rangle=\frac{N_{\alpha}}{2}{\rm
sin}\theta_{\alpha}{\rm sin}\varphi_{\alpha},$ $\langle
S_{\alpha z}\rangle=\frac{N_{\alpha}}{2}{\rm cos}\theta_{\alpha},$ $(\alpha=a,b)$.

For $g_{e}<g_{s}-g_{d}$, $E_{MF}$ is minimal when
$\theta_{a}=0$ while $\theta_{b}=\pi$ or $\theta_{a}=\pi$ while $\theta_{b}=0$, that is, the two
spins are antiparallel and are along $z$ axis.

For
$g_{e}>g_{s}-g_{d}$, $E_{MF}$ is minimal when $\theta_{a}=\theta_{b}=\pi/2$ while $\varphi_{b}=\varphi_{a}+\pi$, that is, the two spins are antiparallel and are on  $x-y$ plane.

One can also use the so-called fidelity susceptibility to
analyze the QPT \cite{fs1,fs2}. For a Hamiltonian ${\cal H}(\eta)={\cal
H}_{0}+\eta {\cal H}_1$, where $\eta$ is a driving parameter, the
ground state fidelity is defined as
$F=|\langle\psi_{0}(\eta+\delta\eta)|\psi_{0}(\eta)\rangle|$. For a nondegenerate ground state,  the fidelity  susceptibility is $
\chi(\eta) \equiv -\lim_{\delta\eta\rightarrow 0}\frac{2 {\rm
ln}F}{\delta\eta^{2}}=\sum_{n\neq0}\frac{|\langle\psi_{n}(\eta)|{\cal
H}_{\alpha}|\psi_{0}(\eta)\rangle|^{2}}{[E_{n}(\eta)-E_{0}(\eta)]^{2}}$,
where $|\psi_{0}(\eta)\rangle$ and $|\psi_{n}(\eta)\rangle$ are the
ground and excited states of ${\cal H}(\eta)$, respectively. In our Hamiltonian, ${\cal H}_{0}=J_zS_{az}S_{bz}$,
${\cal H}_{1}=J_z(S_{ax}S_{bx}+S_{ay}S_{by})$, $\eta \equiv J_{\perp}/J_z=g_e/(g_s-g_d) $.

\begin{figure}
\begin{center}
\includegraphics[height=2in,width=3.2in]{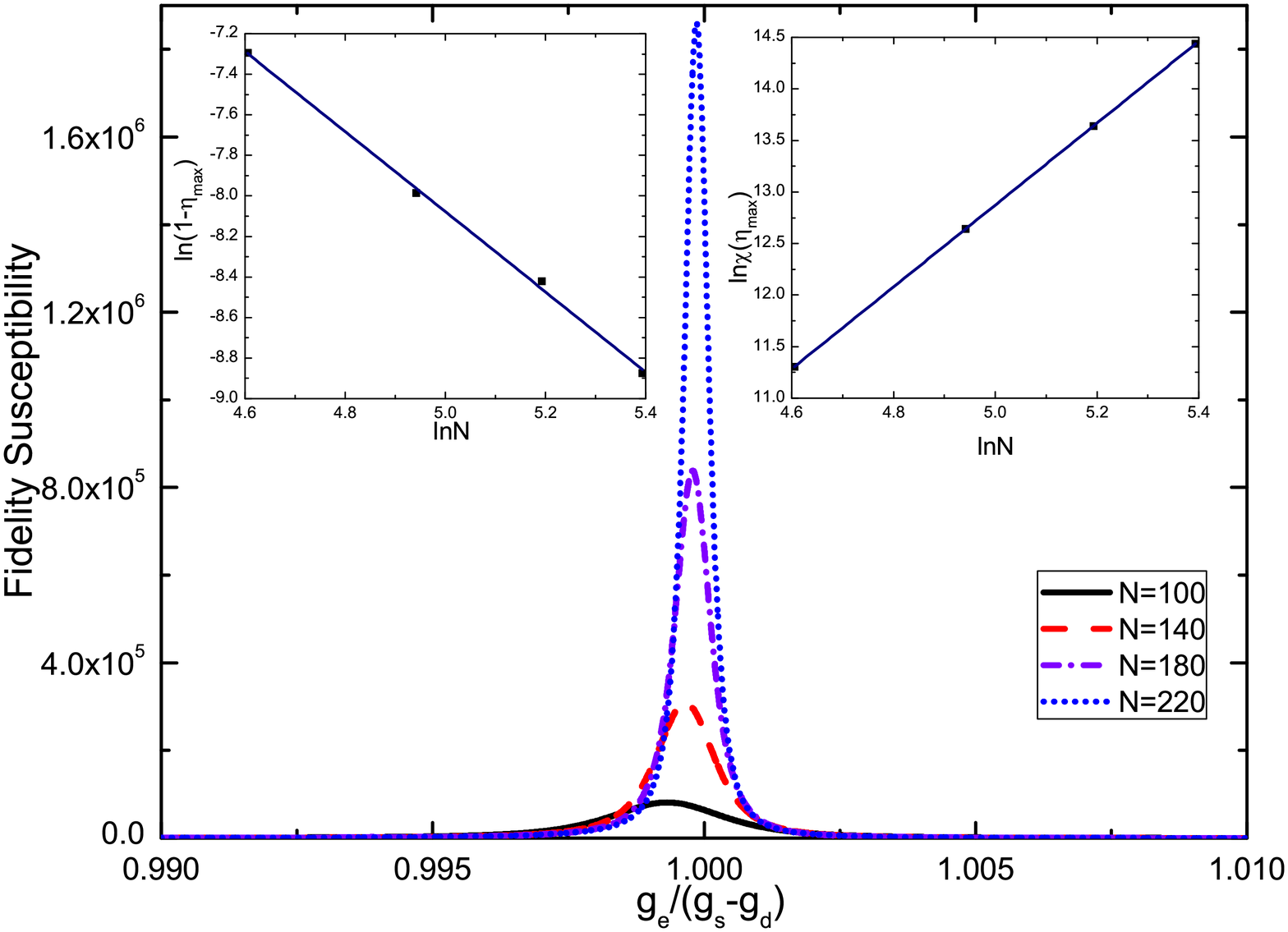}
\caption{Fidelity susceptibility $\chi$ as a function of
$\eta \equiv g_{e}/(g_{s}-g_{d})$ for $N_{a}=N_{b}=N$. The left inset shows how   $\eta_{max}$ varies with   $N$ and the right inset shows how
$\chi(\eta_{max})$ varies with  $N$. \label{fspic}}
\end{center}
\end{figure}

We calculate the fidelity
susceptibility for the case of  $N_{a}=N_{b}$, in which the ground state is non-degenerate. In Fig.\ref{fspic}, $\eta_{max}$ is the position
where the fidelity susceptibility is maximal, $\eta_{max}\sim
1-5.3349 N^{-1.9388}$, the fidelity susceptibility becomes sharper
in the vicinity of $\eta_{max}$ when the number of atoms increases,
$\chi\sim N^{3.971}$. The appearance of the peak implies profound
change of the ground state at $\eta_{max}$. In the thermodynamic
limit, $\eta_{max}$ approaches the maximum $1$, where QPT takes place.

The result of the fidelity susceptibility is very much consistent with the previous result of the entanglement entropy~\cite{shi}. First,  the maxima of both quantities are located at $\eta=1$. Second, the trends in approaching of the maxima are also consistent. The larger the particle number $N$, the quicker both quantities approach the maxima. Third, both quantities display asymmetries between $\eta >1$ and $\eta<1$, with the approaches of both quantities to maxima being quicker on the side of $\eta<1$.

\section{Elementary excitations of the orbital wave functions}

Now we show how the spectra of the  elementary excitations of the orbital wave functions vary with the many-body ground state.  These single-particle wave functions are governed by the the generalized time-dependent Gross-Pitaevskii-like equations, which are determined by minimizing the energy, i.e. the expectation of the Hamiltonian under the many-body ground state, as a functional of the orbital wave functions~\cite{shi}. Previous calculations have been focused on the many-body singlet ground state~\cite{shi3}, here we make a more  general consideration.

From the Hamiltonian, we obtain the Gross-Pitaevskii equations
\begin{equation}
\begin{array}{l}
\displaystyle \langle N_{\alpha\sigma}\rangle i\hbar\frac{\partial}{\partial
t}\phi_{\alpha \sigma} =\{ \langle
N_{\alpha \sigma}\rangle[-\frac{\hbar^{2}}{2m_{\alpha}}
\nabla^{2}+U_{\alpha\sigma}(\textbf{r})] \\ \displaystyle
+\langle
N_{\alpha\sigma}^{2}-N_{\alpha\sigma}\rangle
g_{\alpha}|\phi_{\alpha\sigma}|^{2}
+\langle N_{\alpha\sigma}N_{\alpha\bar{\sigma}}\rangle
g_{\alpha}|\phi_{\alpha \bar{\sigma}}|^{2} \\  \displaystyle  +\langle
N_{\alpha\sigma}N_{\beta\sigma}\rangle
g_{s}|\phi_{\beta\sigma}|^{2}
+\langle N_{\alpha\sigma}N_{\beta\bar{\sigma}}\rangle
g_{d}|\phi_{\beta\bar{\sigma}}|^{2}\} \phi_{\alpha\sigma} \\  \displaystyle     + g_{e}\langle
\alpha^{\dagger}_{\sigma}\alpha_{\bar{\sigma}}
\beta^{\dagger}_{\bar{\sigma}}\beta_{\sigma}\rangle\phi^{*}_{\beta\bar{\sigma}}
\phi_{\beta\sigma}\phi_{\alpha\bar{\sigma}},
\end{array}\label{gpequation}
\end{equation}
where  $\beta \neq \alpha$, $\bar{\sigma} \neq \sigma$,
$\langle O \rangle$  represents the expectation value  of the
operator $O$  in the many-body ground state.

The particle number operators whose expectation values appear in the  Gross-Pitaevskii-like equations can be represented in terms of collective spins of the two species, because of the relations $
N_{\alpha\sigma} = N_{\alpha}/2 + \eta_{\sigma} S_{\alpha z},$
with  $\eta_{\uparrow} =1 $ and  $\eta_{\downarrow} =-1$, and $\langle
\alpha^{\dagger}_{\sigma}\alpha_{\bar{\sigma}}
\beta^{\dagger}_{\bar{\sigma}}\beta_{\sigma}\rangle = \langle S_{ax}S_{bx}+S_{ay}S_{by} \rangle$.  Therefore,  $\langle N_{\alpha\sigma}\rangle = N_{\alpha}/2+ \eta_{\sigma}\langle S_{\alpha z}\rangle$, $\langle N_{\alpha\sigma}^{2}\rangle = N_{\alpha}^2/4 + \eta_{\sigma} N_{\alpha}\langle S_{\alpha z} \rangle + \langle S_{\alpha z}^2 \rangle $, $\langle N_{\alpha\sigma}N_{\alpha\bar{\sigma}}\rangle = N_{\alpha}^2/4  - \langle S_{\alpha z}^2 \rangle $, $ \langle
N_{a\sigma}N_{b \sigma}\rangle = (N_aN_b)/4+ \langle S_{a z}S_{b z}\rangle  + \eta_{\sigma}(N_a/2)\langle S_{b z}\rangle+ \eta_{\sigma}(N_b/2)\langle S_{a z}\rangle $, $\langle N_{\alpha\sigma}N_{\beta\bar{\sigma}}\rangle = (N_aN_b)/4- \langle S_{\alpha z}S_{\beta z}\rangle + \eta_{\bar{\sigma}}(N_{\alpha}/2)\langle S_{\beta z}\rangle  + \eta_{\sigma} (N_{\beta}/2) \langle S_{\alpha z}\rangle$. In the special case of many-body singlet ground state, we have $N_a=N_b=N$ and $\langle S_{az}\rangle = \langle S_{bz}\rangle =\langle S_z \rangle =0$, leading the special form of Gross-Pitaevskii-like equation considered previously.

With these spin quantities as the coefficients, the Gross-Pitaevskii-like equations reflect coupling between spin and orbital degrees of freedom. As the ground state varies with the parameters, so are the expectation values of these spin  quantities.

For simplicity, we focus on the uniform case in absence of external field $U=0$, the lowest energy wave functions are thus $\phi^0_{\alpha\sigma}= \frac{e^{-i\mu_{\alpha\sigma}t/\hbar}}{\sqrt{\Omega}}$, where $\mu_{\alpha\sigma}$ is the chemical potential determined by substituting $\phi^0_{\alpha\sigma}$ in (\ref{gpequation}),
$ \langle N_{\alpha\sigma}\rangle \mu_{\alpha \sigma}  = \langle
N_{\alpha\sigma}^{2}-N_{\alpha\sigma}\rangle
\frac{g_{\alpha}}{\Omega}
+\langle N_{\alpha\sigma}N_{\alpha\bar{\sigma}}\rangle
\frac{g_{\alpha}}{\Omega}        +\langle
N_{\alpha\sigma}N_{\beta\sigma}\rangle
\frac{g_{s}}{\Omega}
+\langle N_{\alpha\sigma}N_{\beta\bar{\sigma}}\rangle
\frac{g_{d}}{\Omega}     +
\langle \alpha^{\dagger}_{\sigma}\alpha_{\bar{\sigma}}
\beta^{\dagger}_{\bar{\sigma}}\beta_{\sigma}\rangle
\frac{g_{e}}{\Omega}.$

Now we consider elementary excitation, i.e. deviation $\delta\phi_{\alpha\sigma} \equiv  \frac{e^{-i\mu_{\alpha\sigma}t/\hbar}}{\sqrt{\Omega}}
[u_{\alpha\sigma}e^{i(\textbf{q}\cdot\textbf{r}-\omega
t)}-v^{*}_{\alpha\sigma}e^{-i(\textbf{q}\cdot\textbf{r}-\omega
t)}]$ from $\phi_{\alpha\sigma}^0$.
Then the Gross-Petaevskii-like equation yields the Bogoliubov-de Gennes-like equations
\begin{equation}
\begin{array}{l}
\displaystyle \langle N_{\alpha\sigma}\rangle (\hbar\omega - E^{\alpha}_{q})u_{\alpha \sigma}
=  \\\displaystyle  \langle
N_{\alpha\sigma}^{2}-N_{\alpha\sigma}\rangle
\frac{g_{\alpha}}{\Omega} (u_{\alpha\sigma} - v_{\alpha\sigma} )
+\langle N_{\alpha\sigma}N_{\alpha\bar{\sigma}}\rangle
\frac{g_{\alpha}}{\Omega} ( u_{\alpha\bar{\sigma} } -v_{\alpha\bar{\sigma} } )  \\ \displaystyle
+\langle
N_{a\sigma}N_{b\sigma}\rangle
\frac{g_{s}}{\Omega} (u_{\beta\sigma } -v_{\beta\sigma} )
+\langle N_{\alpha\sigma}N_{\beta\bar{\sigma}}\rangle
\frac{g_{d}}{\Omega} ( u_{\beta\bar{\sigma}} -v_{\beta\bar{\sigma} } ) \\  \displaystyle  +
\langle \alpha^{\dagger}_{\sigma}\alpha_{\bar{\sigma}}
\beta^{\dagger}_{\bar{\sigma}}\beta_{\sigma}\rangle
\frac{g_{e}}{\Omega} (u_{\beta\sigma} + u_{\alpha\bar{\sigma}} -v_{\beta\bar{\sigma} } -u_{\alpha\sigma} ),
\end{array} \label{uv1}
\end{equation}

\begin{equation}
\begin{array}{l}
\displaystyle \langle N_{\alpha\sigma}\rangle (-\hbar\omega - E^{\alpha}_{q})v_{\alpha \sigma}
=  \\ \displaystyle  \langle
N_{\alpha\sigma}^{2}-N_{\alpha\sigma}\rangle
\frac{g_{\alpha}}{\Omega} (v_{\alpha\sigma} - u_{\alpha\sigma})
+\langle N_{\alpha\sigma}N_{\alpha\bar{\sigma}}\rangle
\frac{g_{\alpha}}{\Omega} (v_{\alpha\bar{\sigma} } -u_{\alpha\bar{\sigma} } )  \\ \displaystyle
+\langle
N_{a\sigma}N_{b\sigma}\rangle
\frac{g_{s}}{\Omega} ( v_{\beta\sigma } -u_{\beta\sigma} )
+\langle N_{\alpha\sigma}N_{\beta\bar{\sigma}}\rangle
\frac{g_{d}}{\Omega} ( v_{\beta\bar{\sigma}} -u_{\beta\bar{\sigma} } ) \\  \displaystyle  +
\langle \alpha^{\dagger}_{\sigma}\alpha_{\bar{\sigma}}
\beta^{\dagger}_{\bar{\sigma}}\beta_{\sigma}\rangle
\frac{g_{e}}{\Omega} (v_{\beta\sigma} + v_{\alpha\bar{\sigma}} -u_{\beta\bar{\sigma} } -v_{\alpha\sigma}).
\end{array} \label{uv2}
\end{equation}
where
$E^{\alpha}_{q}=\frac{\hbar^{2}q^{2}}{2m_{\alpha}}$.

For simplicity, in the following we focus on the ground states with $N_a=N_b=N$,  total $S_z=0$, and thus $\langle S_{\alpha z}\rangle =0$, $\langle S_{\alpha z}^2\rangle =-\langle S_{az}S_{bz}\rangle$. Therefore $\langle N_{\alpha\sigma}\rangle = N/2$, $\langle N_{\alpha\sigma}^{2}\rangle = \langle N_{\alpha\sigma}N_{\beta\bar{\sigma}}\rangle = N^2/4- \langle S_{az}S_{bz}\rangle$, $\langle N_{\alpha\sigma}N_{\alpha\bar{\sigma}}\rangle =\langle
N_{a\sigma}N_{b \sigma}\rangle = N^2/4+\langle S_{az}S_{bz}\rangle$. Hence (\ref{uv1}) and (\ref{uv2}) can be written as
\begin{widetext}
\begin{equation}
\begin{array}{rl}
\displaystyle
&(\hbar\omega-E^{\alpha}_{q})u_{\alpha\sigma}=
\frac{1}{2}\rho g_{\alpha}(1+\lambda_z)(u_{\alpha\sigma}-v_{\alpha\sigma})
+\frac{1}{2}\rho
g_{\alpha}(1-\lambda_z)(u_{\alpha\bar{\sigma}}-v_{\alpha\bar{\sigma}}) \\
\displaystyle
&+\frac{1}{2}\rho g_{s}(1-\lambda_z)(u_{\beta\sigma}-v_{\beta\sigma})
+\frac{1}{2}\rho
g_{d}(1+\lambda_z)(u_{\beta\bar{\sigma}}-v_{\beta\bar{\sigma}})
-\frac{1}{2}\rho g_{e}\lambda_{\perp}(u_{\beta\sigma}+u_{\alpha\bar{\sigma}}-
v_{\beta\bar{\sigma}}-u_{\alpha\sigma}),\\
\end{array}
\label{excitationequation1}
\end{equation}
\begin{equation}
\begin{array}{rl}
\displaystyle
&(-\hbar\omega-E^{\alpha}_{q})v_{\alpha\sigma}
=\frac{1}{2}\rho g_{\alpha}(1+\lambda_z)(v_{\alpha\sigma}-u_{\alpha\sigma})
+\frac{1}{2}\rho
g_{\alpha}(1-\lambda_z)(v_{\alpha\bar{\sigma}}-u_{\alpha\bar{\sigma}})
\\
\displaystyle  & +\frac{1}{2}\rho g_{s}(1-\lambda_z)(v_{\beta\sigma}-u_{\beta\sigma})+\frac{1}{2}\rho
g_{d}(1+\lambda_z)(v_{\beta\bar{\sigma}}-u_{\beta\bar{\sigma}})
-\frac{1}{2}\rho
g_{e}\lambda_{\perp}(v_{\beta\sigma}+v_{\alpha\bar{\sigma}}
-u_{\beta\bar{\sigma}}-v_{\alpha\sigma})=0,
\end{array}
\label{excitationequation2}
\end{equation}
\end{widetext}
where $\rho=N/\Omega$, $\lambda_z \equiv -\langle
S_{az}S_{bz}\rangle/(N^2/4),$  $\lambda_{\perp} \equiv -\langle
S_{ax}S_{bx}+S_{ay}S_{by}\rangle/(N^2/4)$, and we have set $\langle
N_{\alpha\sigma}^{2}-N_{\alpha\sigma}\rangle/\langle N_{\alpha\sigma}\rangle \approx \langle
N_{\alpha\sigma}^{2}\rangle/\langle N_{\alpha\sigma}\rangle$.

It can be found that there are four energy spectra of elementary excitations,\\
\begin{eqnarray}
&\hbar\omega_{1,2}=
\frac{1}{\sqrt{2}}\sqrt{\Gamma_{1}\mp\sqrt{(\Gamma_{1})^{2}+\Gamma_{2}}},\\
&\hbar\omega_{3,4}=
\frac{1}{\sqrt{2}}\sqrt{\Gamma_{3}\mp\sqrt{(\Gamma_{3})^{2}+\Gamma_{4}}},
\label{spectrums}
\end{eqnarray}
where $-$'s are for $\hbar \omega_1$ and $\hbar \omega_3,$ respectively, $+$'s are for $\hbar \omega_2$ and $\hbar \omega_4,$ respectively,
$\Gamma_{1}\equiv E^{a}_{q}(E^{a}_{q}+2\rho g_{a})
+E^{b}_{q}(E^{b}_{q}+2\rho g_{b}),$ $
\Gamma_{2}\equiv 4E^{a}_{q}E^{b}_{q}\{\rho^2 [g_{s}+g_{d}
-(g_{s}-g_{d})\lambda_z-g_{e}\lambda_{\perp}]^{2}
-(E^{a}_{q}+2\rho g_{a})(E^{b}_{q}+2\rho g_{b})\},$ $
\Gamma_{3} \equiv E^{a}_{q}[E^{a}_{q}+2\rho(g_{a}\lambda_z+g_{e}\lambda_{\perp})]
+E^{b}_{q}[E^{b}_{q}+2\rho(g_{b}\lambda_z+g_{e}\lambda_{\perp})]
+2\rho^{2} g_{e}\lambda_{\perp}[(g_{a}+g_{b}+g_{s}+g_{d})\lambda_z+g_{e}
\lambda_{\perp}+g_{d}-g_{s}],$ $
\Gamma_{4} \equiv 4[E^{a}_{q}E^{b}_{q}+\rho
g_{e}\lambda_{\perp}(E^{a}_{q}+E^{b}_{q})]
\{\rho^{2}[g_{s}-g_{d}-(g_{s}+g_{d})\lambda_z]^{2}
-[E^{a}_{q}+\rho(2g_{a}\lambda_z+g_{e}\lambda_{\perp})]
[E^{b}_{q}+\rho(2g_{b}\lambda_z+g_{e}\lambda_{\perp})]
\}$.
If $g_{e}=0$, the Hamiltonian would have four $U(1)$ symmetries
which result in four gapless energy spectra. Now that  $g_{e}\neq 0$,
there are only three $U(1)$ symmetries, three energy spectra are
gapless, and the fourth energy spectra opens a gap~\cite{shi3}. Indeed, when $q=0$ and $g_{e}\neq 0$, $\hbar\omega_{1,2,3}=0$, $\hbar\omega_{4}\neq 0$.

We now study the behavior of the energy gap  $\Delta$ of the forth elementary excitation spectrum as  a function of $g_e$ and $g_s-g_d$. It is easy to write down
\begin{equation}
\begin{array}{rl}
\Delta=&\frac{1}{\sqrt{2}}\sqrt{\Gamma_{3}+
\sqrt{\Gamma_{3}^{2}+\Gamma_{4}}}\\
=& \rho
\sqrt{2g_{e}\lambda_{\perp}[
(g_{a}+g_{b}+g_{s}+g_{d})\lambda_z+g_{e}\lambda_{\perp}+g_{d}-g_{s}]}
\end{array} \label{delta}
\end{equation}

Obviously $\Delta \neq 0$ only if $g_e \neq 0$. Hence interspecies spin exchange is necessary for opening a gap in an orbital excitation.

We have  numerical  calculated $\lambda_z$ and $\lambda_{\perp}$ for various values of  $g_e$ and $g_s-g_d$, and then obtained $\Delta$ by using (\ref{delta}). It is found that when $N\rightarrow\infty$,
\begin{equation}
\begin{array}{rl}
g_{e}<g_{s}-g_{d}: &\lambda_z\rightarrow 1, \mbox{ } \lambda_{\perp}\rightarrow0,  \mbox{ } {\rm hence} \mbox{ } \Delta \rightarrow 0,\\
g_{e}=g_{s}-g_{d}: & \lambda_z \rightarrow 1/3, \mbox{ } \lambda_{\perp}\rightarrow 2/3, \\
g_{e}>g_{s}-g_{d}: &\lambda_z \rightarrow0, \mbox{ } \lambda_{\perp}\rightarrow1.
\end{array} \label{xx}
\end{equation}
It can be seen that the numerical result at $g_{e}=g_{s}-g_{d}$ is the same as the exact result.
$\lambda_{\perp}\rightarrow0$  implies that the effect of spin exchange diminishes, consequently $\omega_{3}\rightarrow\omega_{1}$,
$\omega_{4}\rightarrow\omega_{2}$, the excitations become two doubly degenerate ones. Note that this is under the condition that each scattering strength is symmetric for the two pseudospin states of each species.

\begin{figure}
\begin{center}
\includegraphics[height=2in,width=3.2in]{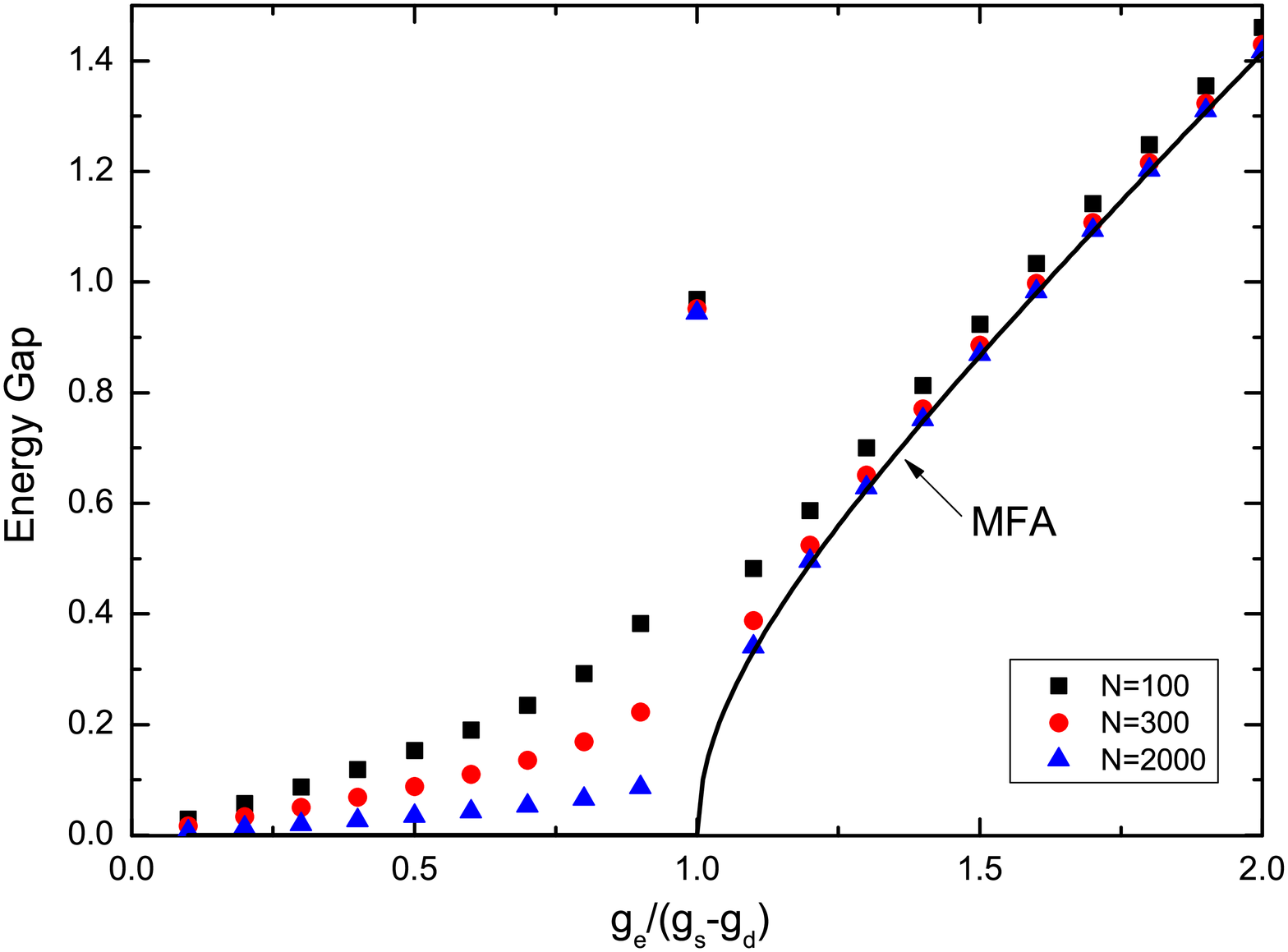}\end{center}
\caption{\label{gap} The energy gap, scaled by
$\sqrt{2}\rho(g_{s}-g_{d})$, as a function of
$g_{e}/(g_{s}-g_{d})$, for different values of N. Points
represent numerical results, while the solid line is from the mean-field disentangled ground state. The parameter values used  satisfy
$(g_{a}+g_{b}+g_{s}+g_{d})/(g_{s}-g_{d})=5.0$ }
\end{figure}

As shown in Fig. \ref{gap}, the larger the  numbers of atoms of
the two species, the closer is the gap  to  that under the disentangled ansatz, except
in the vicinity the critical point,  where the gap varies rapidly with $g_e/(g_{s}-g_{d})$.  At critical point, as the atom numbers increase, the gap quickly saturates to a non-zero value.

Under a disentangled mean-field ground state, one has
$\lambda_z=1$ and  $\lambda_{\perp}=0$ for $g_{e}<g_{s}-g_{d}$, and
$\lambda_z=0$ and $\lambda_{\perp}=1$ for $g_{e}>g_{s}-g_{d}$.  Therefore, far away from the critical
point $g_{e}=g_{s}-g_{d}$, $\lambda_z$ and $\lambda_{\perp}$, and thus the actual elementary excitations, are close to those under the disentangled mean-field states. But the disentangled ansatz clearly fails in the vicinity of the critical point $g_{e}=g_{s}-g_{d}$, as it tells that $\lambda_z$ and $\lambda_{\perp}$ are arbitrary non-negative values satisfying
$\lambda+\lambda_{\perp}=1$.

\section{Summary}

A mixture of two pseudospin-$\frac{1}{2}$ Bose gases with interspecies spin exchange displays interesting interplay between spin and orbital degrees of freedom.
The many-body Hamiltonian is simplified to an anisotropic Heisenberg coupling between the two collective spins of the two species, hence the   particle numbers and correlations and fluctuations are equivalent to the corresponding quantities of the collective spins. These quantities enter the
general Gross-Pitaevskii-like equations governing the four orbital wave functions, in which BEC occurs.  Consequently, the elementary excitations of the orbital wave functions depend on the nature of collective spins in the many-body ground state, and thus serve as probes of entanglement and quantum phase transitions in the latter. Especially, we have shown that the gap of one of the excitations peaks at the antiferromagnetic isotropic parameter point of the effective Heisenberg coupling, which is critical point of quantum phase transition, where the interspecies entanglement and fidelity susceptibility also peak. These properties should be carried over to a spin-1 mixture~\cite{shi5}.

\acknowledgments

We thank Li Ge and  Shi-Jian Gu for useful discussion.
This work was supported by the National Science Foundation of China (Grant No. 11074048), the Shuguang Project (Grant No. 07S402) and the Ministry of Science and Technology of China (Grant No. 2009CB929204).

\end{document}